\documentclass[a4paper,aps,pre,twocolumn,floatfix,showpacs,superscriptaddress]{revtex4-1}

\usepackage[pdftex]{graphicx}
\usepackage{latexsym,amsmath,verbatim}
\usepackage{color}
\usepackage{hyperref}
\usepackage{xspace}
\usepackage{lipsum}

\newcommand{\av}[1]{\left\langle {#1} \right\rangle}
\newcommand{\hi}{Hirsch index\xspace}
\newcommand{\h}[2]{h_{#1}^{(#2)}}

\begin{document}

\title{Topological structure and the $H$-index in complex networks}

\author{Romualdo Pastor-Satorras}

\affiliation{Departament de F\'{\i}sica, Universitat Polit\`ecnica de
  Catalunya, Campus Nord B4, 08034 Barcelona, Spain}

\author{Claudio Castellano}

\affiliation{Istituto dei Sistemi Complessi (ISC-CNR), Via dei Taurini
  19, I-00185 Roma, Italy}

\affiliation{Dipartimento di Fisica, ``Sapienza'' Universit\`a di Roma,
  P.le A. Moro 2, I-00185 Roma, Italy}

\begin{abstract}
  The generalized $H(n)$ \hi of order $n$ has been recently introduced
  and shown to interpolate between the degree and the $K$-core
  centrality in networks. We provide a detailed analytic
  characterization of the properties of sets of nodes having the same
  $H(n)$, within the annealed network approximation. The connection
  between the Hirsch indices and the degree is highlighted. Numerical
  tests in synthetic uncorrelated networks and real-world correlated
  ones validate the findings.  We also test the use
  of the \hi for the identification of influential spreaders in networks,
finding that it is in general outperformed by the recently introduced
Non-Backtracking centrality.
\end{abstract}


\maketitle

\section{Introduction}
\label{sec:introduction}

Many topological properties have been proposed and measured to
characterize complex networks~\cite{Newman10}.  Among them, centrality
measures (such as degree, betweenness \cite{freeman77} or eigenvector
\cite{Bonacich72} centrality) aim at quantifying the relative importance
of individual vertices in the overall topology; they are often related
to the behavior of processes unfolding on the complex network structure,
such as spreading or diffusion~\cite{Klemm2012,Goltsev12,Lu16}.
Prominent in this context is the $K$-core decomposition, a recursive
pruning procedure~\cite{Seidman1983269} which iteratively peels off
nodes from the network ($K$-shells), leaving subsets ($K$-cores) which
are increasingly dense and mutually interconnected.  The $K$-core
decomposition proceeds as follows: Starting with the full graph, nodes
with degree $q=1$ are removed, repeating this operation iteratively
until only nodes with degree $q \ge 2$ remain. The removed nodes
constitute the $K=1$-shell, and those that remain are the
$K=2$-core. Next, all nodes with degree $q=2$ are iteratively removed,
yielding the $K=2$-shell and the $K=3$-core (remaining network). The
process is repeated until one more iteration removes all nodes.  The
coreness $k_i$ of a node is the index of the $K$-shell to which node $i$
belongs, and it has been argued that the set of nodes with maximum
coreness plays an important role in epidemic spreading
\cite{Castellano2012}.

Recently, a new set of centrality measures has been proposed
\cite{Lu2016}, based on the concept of the Hirsch $H$-index introduced
to quantify research impact of scientists \cite{Hirsch15112005}.  The
$H$-index $h_i$ of node $i$ in a network is defined in analogy to the
Hirsch index for citations: It is the maximum number $h$, such that
there are at least $h$ neighbors of this node with degree larger than or
equal to $h$. In Ref.~\cite{Lu2016}, this concept is generalized to a
hierarchy of $n$-th order Hirsch indices, $H(n)$, using the following
construction: Define an operator $\mathcal{H}(x_1, x_2, \ldots, x_n)$
that is equal to the maximum integer $y$ such that there are at least
$y$ elements in $(x_1, x_2, \ldots, x_n)$ with a value larger than or
equal to $y$.  Defining $\h{i}{0} \equiv q_i$, the degree of node $i$,
the $H$-index, $h_i \equiv \h{i}{1}$, is defined by
\begin{equation}
  \h{i}{1} =
  \mathcal{H}\left( \{\h{j}{0} \}_{j \in \mathcal{V}_i} \right),
  \label{eq:14}
\end{equation}
where $\mathcal{V}_i$ is the set of nearest neighbors of node $i$. The
$n$-th order \hi, $H(n)$, of node $i$, $\h{i}{n}$, is defined by the
iterative relation
\begin{equation}
  \h{i}{n} = \mathcal{H}\left( \{\h{j}{n-1} \}_{j \in \mathcal{V}_i}
  \right). 
  \label{eq:15}
\end{equation}
In Ref.~\cite{Lu2016} it is proved that
\begin{equation}
  \lim_{n\to\infty}  \h{i}{n} = k_i,
\end{equation}
the coreness of node $i$.

This identification allows to make an analogy with the $K$-core
organization of a network, and define a hierarchy of $H(n)$-shells and
$H(n)$-cores. A $H(n)$-shell is the set of nodes with $n$-th order \hi
equal to some value $h$; a $H(n)$-core is defined as the set of all
nodes with $n$-th order \hi larger than or equal to a given value
$h$. The relevance of this organization has been discussed in
Ref.~\cite{Lu2016}, where it is argued that, in some instances, the
$H(n)$ index of a node can be a better predictor of the influence of a
node \cite{Lu16} in epidemic spreading than the degree or the coreness.

Here we study the relation between the different $H(n)$ indices and the
degree $q$ of the corresponding node within the annealed network
approximation \cite{dorogovtsev07:_critic_phenom,Boguna09}. In the case
of uncorrelated networks, we observe that the $n$-th order $H$-index and
the degree of a node are strongly linked. This sort of correlation
extends also, in some cases, to real networks, rife with degree
correlations \cite{assortative}. Moreover, we test the validity of the
\hi as a predictor of spreading influence. We find its performance to be
sometimes slightly better than degree or $K$-core centralities but generally
largely worse than the recently introduced Non-Backtracking (NBT)
centrality~\cite{2014arXiv1401.5093M,Radicchi2016}.

The paper is organized as follows: In Sec.~\ref{sec:structure-hi-cores}
we present a general theoretical description of the relation between
$H(n)$ index and degree, within the annealed network approximation. The
case of uncorrelated networks \cite{Dorogovtsev:2002} is considered in
detailed in Sec.~\ref{sec:uncorr-netw}, where theoretical predictions
are checked against numerical results in random networks. In
Sec.~\ref{sec:correlated-networks} we study the structure of the $H(n)$
shells in real correlated networks, discussing when departures from the
uncorrelated predictions are observed. In Sec.~\ref{sec:test-hi-as} we
present a study of the performance of $H(n)$ index as predictor of
influence in epidemics.  Finally, our conclusions are presented in
Sec.~\ref{sec:conclusions}.

\section{Topological structure of the $H(n)$-shells}
\label{sec:structure-hi-cores}

We study the topological structure of the $H(n)$-shells in terms of the
conditional probability that node with degree $q$ has $H(n)$ index equal
$h$, $P^{(n)}(h|q)$, and the conditional probability that a node with
$H(n)$ index equal $h$ has degree $q$, $P^{(n)}(q|h)$.  These
conditional probabilities are related to the probability $P^{(n)}(h)$
that a randomly chosen node has $H(n)$ index equal to $h$ and to the
degree distribution $P(q)$ by the bayesian relation
\begin{equation}
  P^{(n)}(h) P^{(n)}(q \vert h) = P(q) P^{(n)}(h|q).
  \label{eq:1}
\end{equation}
These probability distributions can be easily estimated within the
annealed network approximation \cite{dorogovtsev07:_critic_phenom}, in
which a network is defined exclusively by its degree distribution
$P(q)$, and the conditional probability $P(q'|q)$ that an edge from a
node of degree $q$ is connected to a node of degree $q'$ \cite{alexei},
being the network random in all other respects.

In the case $n=1$, the cumulative probability $P^{(1)}_c(h|q)$ that a
node of degree $q$ has $H(1)$ index larger than or equal to $h$ is equal
to the probability that it is connected to $h$ or more nodes with degree
larger than or equal to $h$, that is
\begin{equation}
  P^{(1)}_c(h|q) = \sum_{m=h}^{q} \binom{q}{m} [R_{h, q}^{(1)}]^m
  [1-R_{h, q}^{(1)}]^{q-m},
  \label{eq:5}
\end{equation}
where $R_{h, k}^{(1)}$, the probability that an edge from a node of
degree $q$ leads to a node with degree larger than or equal to $h$, is
given by
\begin{equation}
  R_{h, q}^{(1)} = \sum_{q'=h}^\infty P(q'|q).
  \label{eq:3}
\end{equation}
For $n > 1$, the cumulative probability $P^{(n)}_c(h|q)$ that a node of
degree $q$ has a $H(n)$ index larger than or equal to $h$, is equal to
its probability of having $h$ or more neighbors with $H(n-1)$ index
larger than or equal to $h$. Therefore,
\begin{equation}
  P^{(n)}_c(h | q) = \sum_{m=h}^q \binom{q}{m} [R^{(n)}_{h,q}]^m
  [1-R^{(n)}_{h,q}]^{q-m} ,
  \label{eq:19}
\end{equation}
where $R^{(n)}_{h,q}$ is the probability that an edge from a node with
degree $q$ points to a node with $H(n-1)$ index larger than or equal to
$h$.  Within the annealed network approximation, we can write
\begin{equation}
  R^{(n)}_{h,q} =\sum_{h' =h}^\infty \sum_{q'} P(q ' | q) P^{(n-1)}(h | q'),
  \label{eq:20}
\end{equation}
which is constructed considering that the node $q$ is connected to a
node of degree $q'$ with probability $P(q ' | q)$, and that this one has
$H(n-1)$ index equal to $h$ with probability $P^{(n-1)}(h | q')$.
Rearranging the summation in Eq.~(\ref{eq:20}) and inserting the
definition of $P^{(n-1)}_c(h | q)$ from Eq.~(\ref{eq:19}) into it we
obtain
\begin{equation}
  R^{(n)}_{h,q} =\sum_{q'} P(q ' | q) \sum_{m=h}^{q'} \binom{q'}{m}
  [R^{(n-1)}_{h,q'}]^m  [1-R^{(n-1)}_{h,q'}]^{q'-m} ,
  \label{eq:32}
\end{equation}
a direct iterative relation between $R^{(n)}_{h,q}$ and
$R^{(n-1)}_{h,q}$, which can be solved with the initial condition
Eq.~(\ref{eq:3}). Together with Eq.~(\ref{eq:19}), Eq.~(\ref{eq:32})
completely determines the topological structure of the $H(n)$-shells at
all orders $n$.

\section{Uncorrelated networks}
\label{sec:uncorr-netw}

Let us consider in detail the case of uncorrelated networks, with
$P(q' | q) = q' P(q')/\av{q}$ \cite{Dorogovtsev:2002}. In this case,
\begin{equation}
  R_{h,q}^{(1)} = \frac{1}{\av{q}}\sum_{q' = h}^\infty q' P(q') \equiv
  R^{(1)} (h),
\end{equation}
independent of $q$, which greatly simplifies calculations.  Considering
the continuous degree approximation in the interesting case
heterogeneous networks with of a power-law degree distribution
$P(q) = (\gamma - 1) m^{\gamma-1} q^{-\gamma}$ \cite{Barabasi:1999},
where $m$ is the minimum degree in the network, we have
\begin{equation}
  R^{(1)}(h) = \left( \frac{h}{m}\right)^{2-\gamma},
  \label{eq:9}
\end{equation}
a decreasing function of $h$ for $\gamma>2$ (imposed to ensure a finite
average degree). Now, from the cumulated conditional probability
$P^{(1)}_c(h|q)$ we have
$ P^{(1)}(h|q) = P^{(1)}_c(h|q) - P^{(1)}_c(h+1|q)$.  Since $R_h$ is a
slowly (algebraic) decreasing function of $h$, we can write
\begin{equation}
  P^{(1)}(h|q) \simeq \binom{q}{h} [R^{(1)}(h)]^h
  [1-R^{(1)}(h)]^{q-h}.
  \label{eq:17}
\end{equation}
That is, $P^{(1)}(h|q)$ is a strongly peaked distribution, centered
around a peak $\overline{h^{(1)}}(q) = \sum_h h P^{(1)}(h|q)$ given by the
implicit equation
\begin{equation}
  \label{eq:2}
  \overline{h^{(1)}}(q)  \simeq q \; R^{(1)}(\overline{h^{(1)}}(q)).
\end{equation}
Applying Eq.~(\ref{eq:9}), we can identify $\overline{h^{(1)}}(q)$,
that is, the average $H(1)$ index of nodes of degree $q$, as
\begin{equation}
  \label{eq:4}
  \overline{h^{(1)}}(q) \sim q^{1/\alpha_1},\quad  \mathrm{with} \quad
  \alpha_1 =  \gamma-1 .
\end{equation}
From this last expression, we can obtain information on the global
distribution of the \hi in the whole network by using
$ P^{(1)}(h) dh \simeq P(q) dq$.  From here, using $q \sim h^{\gamma-1}$
from Eq.~(\ref{eq:4}), we are led to
\begin{equation}
 P^{(1)}(h) \sim h^{-\gamma_1}, \quad  \mathrm{with} \quad
  \gamma_1 =  (\gamma-1)^2 +1.
\label{eq:12}
\end{equation}
That is, the distribution of Hirsch indices in power-law networks
follows also a power-law form, with an exponent that increases
quadratically with the degree exponent.

For the $n$-th order \hi, using the bayesian relation Eq.~(\ref{eq:1}),
we can write relation Eq.~(\ref{eq:20}) as
\begin{equation}
  \label{eq:22}
  R^{(n+1)}(h) = \frac{1}{\av{q}}\sum_{h' \geq h}  P^{(n)}(h')
  \overline{q^{(n)}}(h'),
\end{equation}
where we have defined
$\overline{q^{(n)}}(h) = \sum_{q} q P^{(n)}(q | h)$ as the average
degree of nodes with $n$-th order \hi equal to $h$. Let us define
analogously $\overline{h^{(n)}}(q) = \sum_hh P^{(n)}(h|q)$ as the
average $H(n)$ index of the nodes of degree $q$. Assuming
$P^{(n)}(h) \sim h^{-\gamma_n}$,
$\overline{q^{(n)}}(h) \sim h^{\alpha_n}$,
$\overline{h^{(n)}}(q) \sim q^{1/\alpha_n}$, in analogy with what is
observed in the $H(1)$ case, and using again the relation between
probability distributions $P^{(n)}(h) dh \sim P(q) dq$, we obtain
$\gamma_n = \alpha_n(\gamma-1) +1$, while using Eq.~(\ref{eq:22}) leads
$R^{(n+1)}(h) \sim h^{-\alpha_n(\gamma-2)}$.  Assuming also that
$P^{(n+1)}(h|q)$ is a peaked function, centered at
$\overline{h^{(n+1)}}(q)$, we have, from Eq.~(\ref{eq:19}),
$\overline{h^{(n+1)}}(q) \sim q R^{(n+1)}(\overline{h^{(n+1)}}(q))$,
from where we obtain
$\overline{h^{(n+1)}}(q) \sim q^{1/(1+\alpha_n(\gamma-2))}$, leading, by
comparison with the form $\overline{q^{(n)}}(h) \sim h^{\alpha_n}$, to
the relation
\begin{equation}
  \alpha_{n+1} = \alpha_n(\gamma-2) + 1.
\label{eq:10}
\end{equation}
With the initial condition $\alpha_1 = \gamma-1$, we can solve
Eq.~(\ref{eq:10}) to find
\begin{equation}
 \left\{ \begin{array}{c}
     \overline{q^{(n)}}(h) \sim h^{\alpha_n}\\
    \\
    \overline{h^{(n)}}(q) \sim
    q^{1/\alpha_n}
  \end{array}\right., \;  \mathrm{with} \;\; \alpha_n =
  \frac{(\gamma-2)^{n+1} -1}{\gamma -3}
  \label{eq:16}
\end{equation}
and
\begin{equation}
  P^{(n)}(h) \sim h^{-\gamma_n} , \;  \mathrm{with} \;\; \gamma_n =
  \frac{(\gamma-1)(\gamma-2)^{n+1} -2}{\gamma-3}.
  \label{eq:25}
\end{equation}

As a validation of this result, we recall that the coreness distribution
corresponds to the limit $n \to \infty$. In this limit, for $\gamma <3$,
we obtain $\alpha_\infty = \frac{1}{3-\gamma}$ and
$\gamma_\infty = \frac{2}{3-\gamma}$, while for $\gamma>3$, both
$\alpha_n$ and $\gamma_n$ diverge, indicating that there is no $K$-core
structure. These findings are in agreement with those obtained
rigorously in Refs.~\cite{Goltsev2006,Dorogovtsev2006}.

\begin{figure}[t]
  \includegraphics[width=\columnwidth]{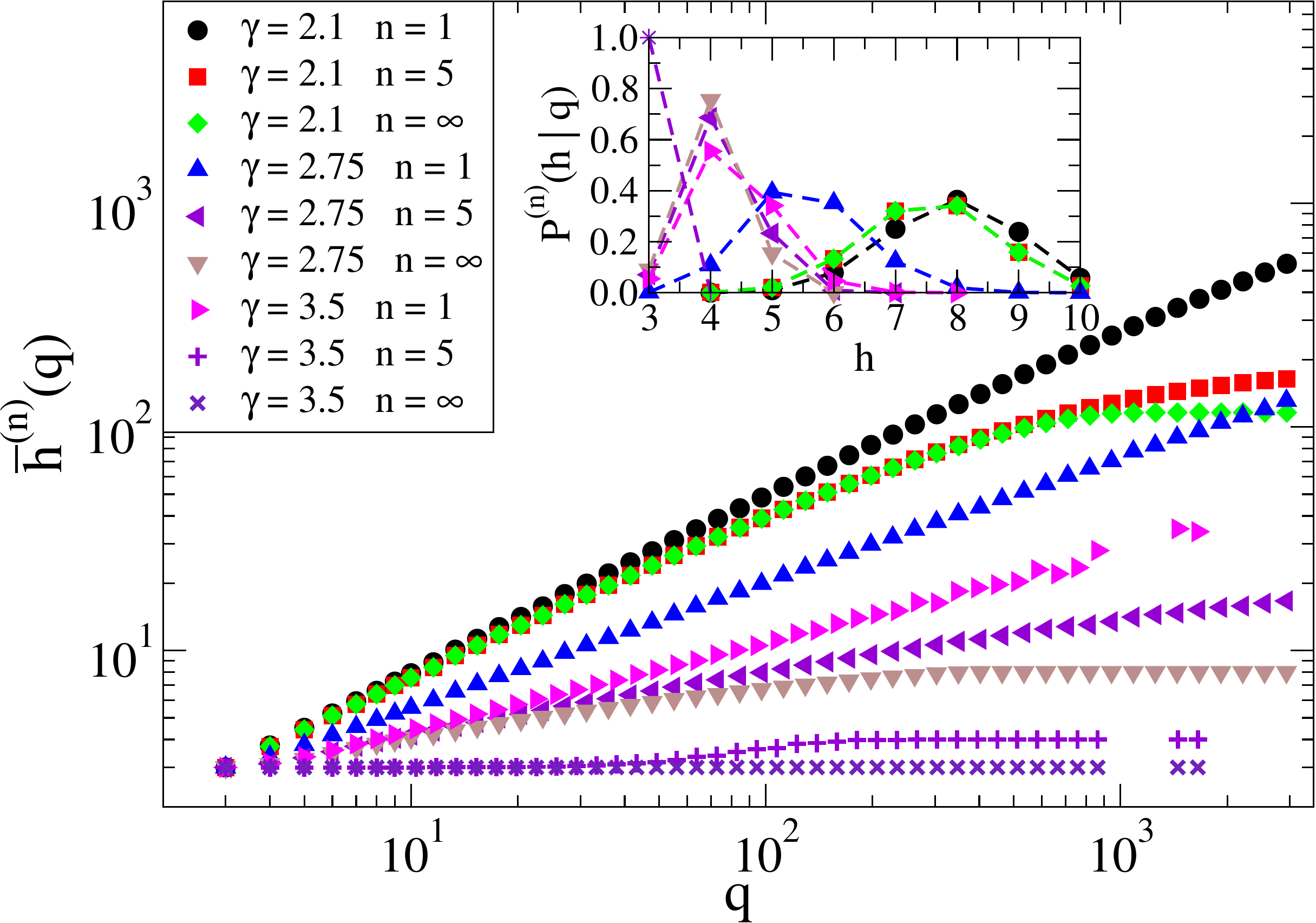}
  \caption{Main plot: Average value of the Hirsch index of order $n$
    versus degree $q$ for nodes of a power-law distributed network of
    degree $\gamma$.  Inset: Conditional probability that a node of
    degree $q=10$ has $n$-th order \hi $h$, for various values of $n$
    and of the degree exponent $\gamma$. The size of the networks is
    $N=10^7$.}
  \label{fig:hn_of_q}
\end{figure}

In order to check the previous results for finite values of $n$, we
perform a numerical analysis of the $H(n)$-shell structure in
uncorrelated scale-free networks, with degree distribution
$P(k) \sim k^{-\gamma}$ and minimum degree $m=3$, generated using the
uncorrelated configuration model (UCM) \cite{Catanzaro05}.

\begin{figure}[t]
  \includegraphics[width=\columnwidth]{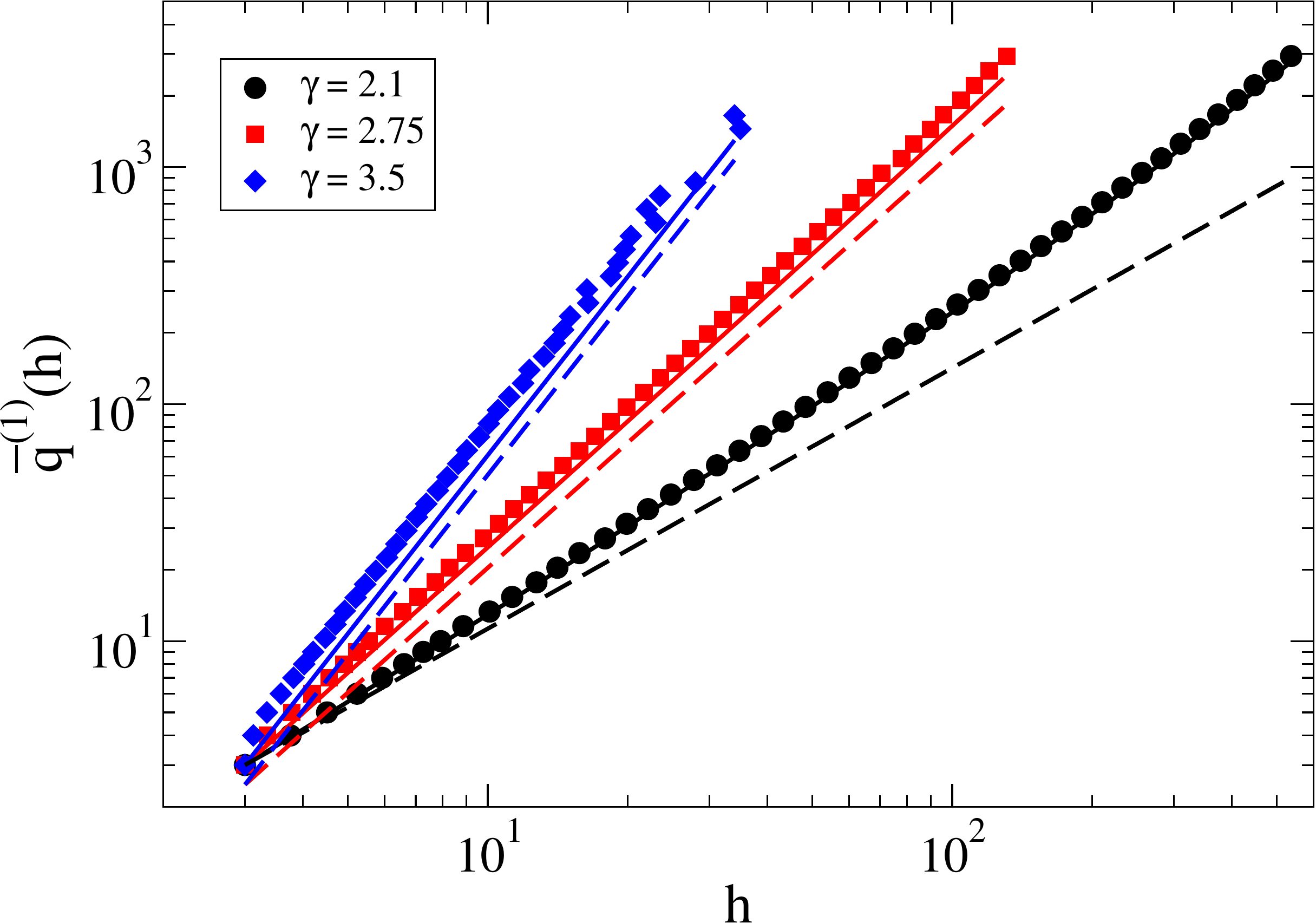}
  \caption{Average value of the degree $q$ versus the \hi of order $n=1$
    of nodes in a power-law distributed network of degree $\gamma$.
    Full lines represent the finite-size form Eq.~(\ref{eq:7}), while
    the dashed lines have the slope of the asymptotic expression
    Eq.~(\ref{eq:4}) (dashed lines have been shifted for clarity). The
    size of the network is $N=10^7$.}
  \label{fig:finite_size}
\end{figure}

In the inset of Fig.~\ref{fig:hn_of_q} we show that the main assumption
leading to the theoretical estimates, namely the peaked form of the
conditional distributions $P^{(n)}(h|q)$, is well satisfied by the
numerical data, the distributions showing a small dispersion relative to
their average value, given by the peak.  In Fig.~\ref{fig:hn_of_q} (main
plot) we plot $\overline{h^{(n)}}$ as a function of $q$ for different
values of the index $n$ and of the degree exponent $\gamma$.  By fitting
the curves in the range $20 \le q \le 200$ to the theoretical prediction
$\overline{h^{(n)}}(q) \sim q^{1/\alpha_n}$ we obtain the values for the
exponent $1/\alpha_n$ reported in Table~\ref{table1}.
\begin{table}[b]
  \begin{ruledtabular}
    \begin{tabular}{||c|c||c|c||}
      $\gamma$ & $n$ & $1/\alpha_n$ (Theory)&  $1/\alpha_n$ (Numerics)\\
      \hline
      \hline
      2.1 & 1 & 0.91 & 0.778 $\pm$ 0.001 \\
      \hline
      2.1 & 5 & 0.90 & 0.673 $\pm$ 0.002 \\
      \hline
      2.75 & 1 & 0.57 & 0.563 $\pm$ 0.001 \\
      \hline
      2.75 & 5 & 0.30 & 0.267 $\pm$ 0.001 \\
      \hline
      3.5 & 1 & 0.40 & 0.404 $\pm$ 0.005 \\
      \hline
      3.5 & 5 & 0.05 & N/A \\
    \end{tabular}
  \end{ruledtabular}

  \caption{Comparison between the values of $1/\alpha$ predicted by
    Eq.~(\ref{eq:16}) and obtained by fitting in Fig.~\ref{fig:hn_of_q}.}
  \label{table1}
\end{table}
As we can observe, the agreement between numerical data and the
theoretical prediction is reasonably good, becoming better for larger
values of $\gamma$. This discrepancy can be attributed to finite-size
effects which, for $n=1$, modify Eq.~(\ref{eq:9}).
For a network of finite size $N$ and correspondingly finite maximum
degree $k_c$, one has instead
\begin{equation}
  \label{eq:6}
  R^{(1)}(h) = \frac{\int_h^{k_c} q^{-\gamma +1}\;dq}{\int_m^{k_c}
    q^{-\gamma +1}\;dq} = \frac{h^{2-\gamma} -
    k_c^{2-\gamma}}{m^{2-\gamma} - k_c^{2-\gamma}} .
\end{equation}
Inserting this expression into Eq.~(\ref{eq:2}), we obtain (assuming
a change of dependency from $\overline{h^{(1)}}(q)$ to
$\overline{q^{(1)}}(h)$)
\begin{equation}
  \label{eq:7}
  \overline{q^{(1)}}(h) \simeq h \frac{m^{2-\gamma} -
    k_c^{2-\gamma}}{h^{2-\gamma} -  k_c^{2-\gamma}} .
\end{equation}
For large $\gamma$ and $N$, the factor $k_c^{2-\gamma}$ is negligible
and we recover the scaling form in Eq.~(\ref{eq:4}),
$\overline{q^{(1)}}(h) \sim h^{\gamma-1}$. For small $\gamma$, however,
one might need a very large $N$ to observe the final asymptotic
regime. This fact is checked in Fig.~\ref{fig:finite_size}, where we
plot $\overline{q^{(1)}}(h)$ for different values of $\gamma$, together
with Eq.~(\ref{eq:7}) and the scaling form Eq.~(\ref{eq:4}). In
Eq.~(\ref{eq:7}), we impose a maximum degree growing with network size
as $k_c(N) = N^{1/\mu}$, with $\mu=2$ for $2 < \gamma \leq 3$ and
$\mu=\gamma-1$ for $\gamma \geq 3$ \cite{mariancutofss}. As we can see,
for $\gamma \geq 2.75$, the finite-size expression and the scaling
form are indistinguishable; that is not the case for $\gamma=2.1$, where
the finite-size form provides a perfect fit to numerical data, while the
asymptotic expression is markedly different.

\begin{figure}[t]
  \includegraphics[width=\columnwidth]{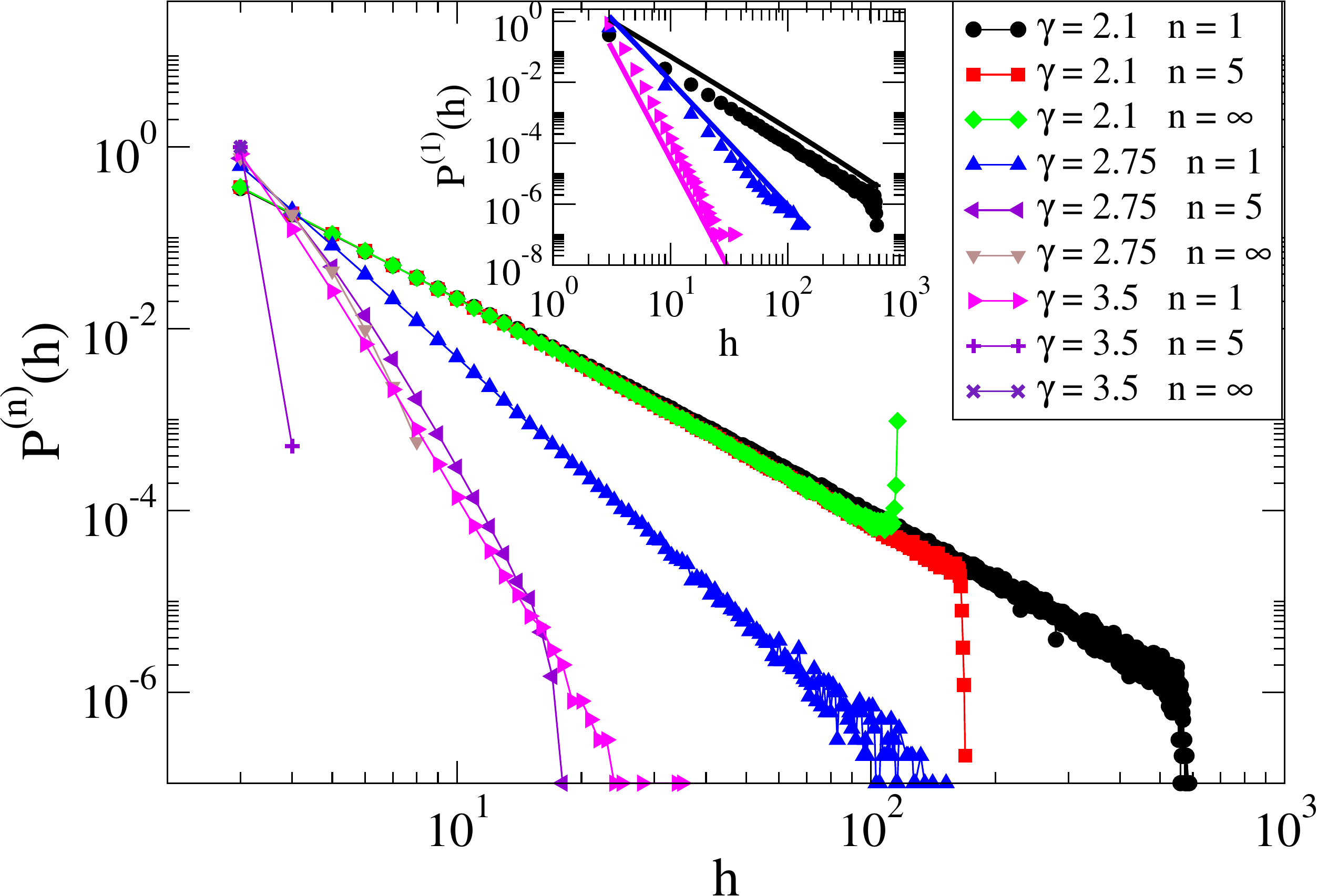}
  \caption{Main plot: Probability density that a node has $n$-th order
    \hi $h$, for various values of $n$ and of the degree exponent
    $\gamma$. Inset: Probability density $P^{(1)}(h)$ versus $h$,
    compared with the finite size scaling form Eq.~(\ref{eq:8}) (full
    lines). The size of the network is $N=10^7$.}
  \label{fig:P_of_h}
\end{figure}

In Fig.~\ref{fig:P_of_h} we plot finally the distribution of $H(n)$
indices for different values of $n$ and $\gamma$. They all show a
long-tailed form, compatible with the prediction
$P^{(n)}(h) \sim h^{-\gamma_n}$, except in the limit $n \to \infty$ for
$\gamma>3$.  The theoretical values of $\gamma_n$ predicted by
Eq.~(\ref{eq:25}) provide a good approximation to the numerical exponents
of the distributions (see Table~\ref{table2}), again with some
discrepancies for small $\gamma$. These can be attributed to the
finite-size effects discussed above. In fact, for $n=1$, considering the
form in Eq.~(\ref{eq:7}) in the calculation of $P^{(1)}(h)$, leads to
the more complex expression
\begin{equation}
  \label{eq:8}
  P^{(1)}(h) \sim h^{-\gamma}\frac{(\gamma-1)k^{2-\gamma} -
    k_c^{2-\gamma}}{(k^{2-\gamma} - k_c^{2-\gamma})^{2-\gamma} },
\end{equation}
which fits remarkably well the numerical data, see inset in
Fig.~\ref{fig:P_of_h}.

\begin{table}[t]
  \begin{ruledtabular}
    \begin{tabular}{||c|c||c|c||}
      $\gamma$ & $n$ &  $\gamma_n$ (Theory) &  $\gamma_n$ (Numerics)\\
      \hline
      \hline
      2.1 & 1 & 2.21 & 2.378 $\pm$ 0.004 \\
      \hline
      2.1 & 5 & 2.22 & 2.494 $\pm$ 0.005 \\
      \hline
      2.75 & 1 & 4.06 & 4.16 $\pm$ 0.01 \\
      \hline
      2.75 & 5 & 6.75 & 7.75 $\pm$ 0.08 \\
      \hline
      3.5 & 1 & 7.25 & 7.57 $\pm$ 0.05 \\
      \hline
      3.5 & 5 & 52.95 & N/A \\
    \end{tabular}
  \end{ruledtabular}

  \caption{Comparison between the values of $\gamma_n$ predicted by
    Eq.~(\ref{eq:25}) and those obtained by fitting in Fig.~\ref{fig:P_of_h}
    in the range $10^{-4} < P^{(n)}(h)< 10^{-2}$.}
\label{table2}
\end{table}

\section{Real world networks}
\label{sec:correlated-networks}

The previous results were obtained in the case of uncorrelated synthetic
networks. To ascertain the effects of correlations and other topological
features, we proceed to compute the average $\overline{h^{(n)}}(q)$ for
different values of $n$ in several real-world heterogeneous correlated
networks. We consider in particular the following real network datasets:
\begin{itemize}
\item Internet AS: Internet map at the autonomous system  level
  \cite{romuvespibook};
\item P2P: Gnutella peer-to-peer file sharing network
  \cite{leskovec_snap_nets};
\item WWW: Notre Dame University Word-Wide Web graph \cite{www99};
\item Polblogs: Network of blogs on US
  politics~\cite{adamic2005political};
\item PGP: Networks of users of the pretty-good-privacy encryption
  algorithm~\cite{PhysRevE.70.056122};
\item Email: Email communication network collected at the Rovira Virgili
  University \cite{PhysRevE.68.065103};
\item Facebook: New Orleans regional Facebook network
  \cite{viswanath2009evolution};
\item Jazz: Network of collaboration among jazz
  musicians~\cite{gleiser2003community} 
\end{itemize}
The main topological properties of these networks are summarized in
Table~\ref{table_topology}.

\begin{table}[b]
  \begin{ruledtabular}
    \begin{tabular}{||c||c|c|c|c|c||c||}
      Network  & $N$ & $\av{q}$ &  $\kappa$ & $r$ & $n_\mathrm{max}$ &$\beta_c$\\
      \hline
      \hline
      Internet AS & 10790 & 4.16 & 61.30 & -0.1938 & 9  &0.0260\\ \hline
      P2P         & 62586 & 4.73 & 1.46  & -0.0926 & 36 &0.1015\\ \hline
      WWW         & 325729& 6.69 & 40.93 & -0.0534 & 187&0.0105\\ \hline
      Polblogs    & 1224  & 27.31& 1.98  & -0.2212 & 18 &0.0160\\ \hline \hline
      PGP         & 10680 & 4.55 & 3.15  &  0.2381 & 14 &0.0590\\ \hline
      Email       & 1133  & 9.62 & 0.94  &  0.0782 & 16 &0.0630\\ \hline
      Facebook    & 63731 & 25.64& 2.43  &  0.1769 & 63 &0.0090\\ \hline
      Jazz        & 198   & 27.70& 0.40  &  0.0202 & 13 &0.0285\\
    \end{tabular}
  \end{ruledtabular}
  \caption{Topological properties of the real networks considered:
    Network size $N$; average degree $\av{q}$; heterogeneity parameter
    $\kappa = \av{q^2}/\av{q}^2 -1$; degree correlations as measured by the 
    Pearson coefficient $r$ \cite{assortative}; maximum Hirsch order
    $n_\mathrm{max}$, leading to the node coreness. The critical point
    $\beta_c$ of SIR processes in the networks, estimated numerically as
    the maximum of the susceptibility \cite{Castellano16},
    is presented in the last row.}
  \label{table_topology}
\end{table}

\begin{figure}[t]
  \includegraphics[width=\columnwidth]{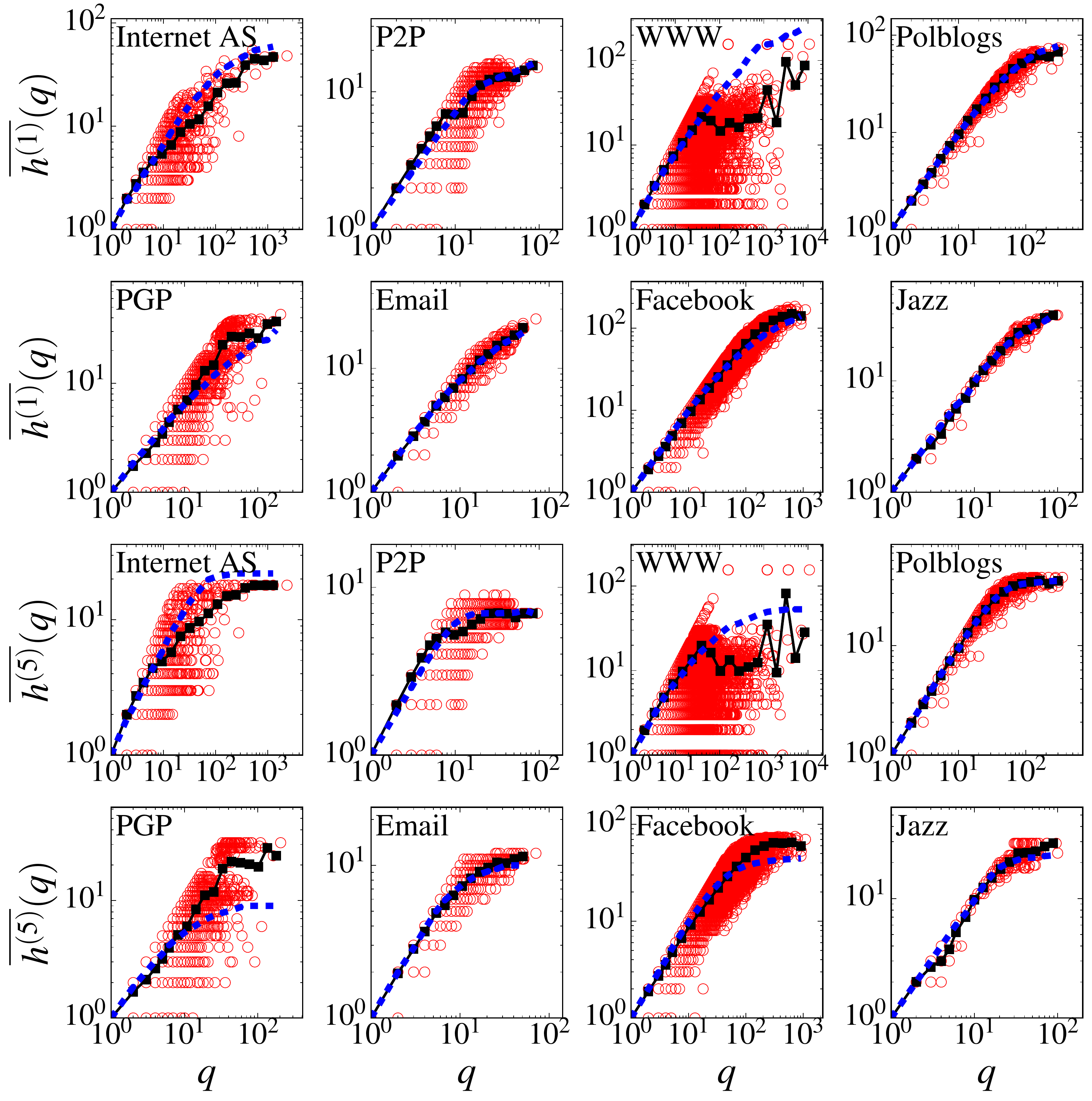}
  \caption{Scatter plot (hollow symbols) of the Hirsch-index of order
    $n=1$ (top plots) and $n=5$ (bottom plots) versus degree $q$ for
    several real-world networks. Full symbols represent the numerical
    averages $\overline{h^{(n)}}(q)$; dashed lines represent the average
    $\overline{h^{(n)}}_\mathrm{ran}(q)$ obtained from randomized
    versions of the networks.}
  \label{fig:real1}
\end{figure}

Figure~\ref{fig:real1} presents a scatter plot of
$\overline{h^{(n)}}(q)$ as a function of $q$ for $n=1$ and $n=5$,
evaluated for these real networks. Full symbols represent the numerical
averages $\overline{h^{(n)}}(q)$, while the dashed lines represent the
averages $\overline{h^{(n)}}_\mathrm{ran}(q)$ computed from randomized
versions of the networks, in which correlations have been washed out by
means of a degree-preserving rewiring procedure \cite{maslov02}. From
this figure we can observe that, in these heterogeneous networks, the
averages $\overline{h^{(n)}}(q)$ follow an approximate algebraic
behavior, in agreement with the theoretical prediction for heterogeneous
networks with a pure power-law degree distribution. The spread of the
$H(n)$ index with respect to its average value is, in general, 
apparently smaller in networks with assortative degree correlations (i.e,
with positive Pearson correlation coefficient, see
Table~\ref{table_topology}), as compared with disassortative networks
(i.e. with negative $r$) \cite{assortative}.
\begin{figure}[t]
  \includegraphics[width=0.9\columnwidth]{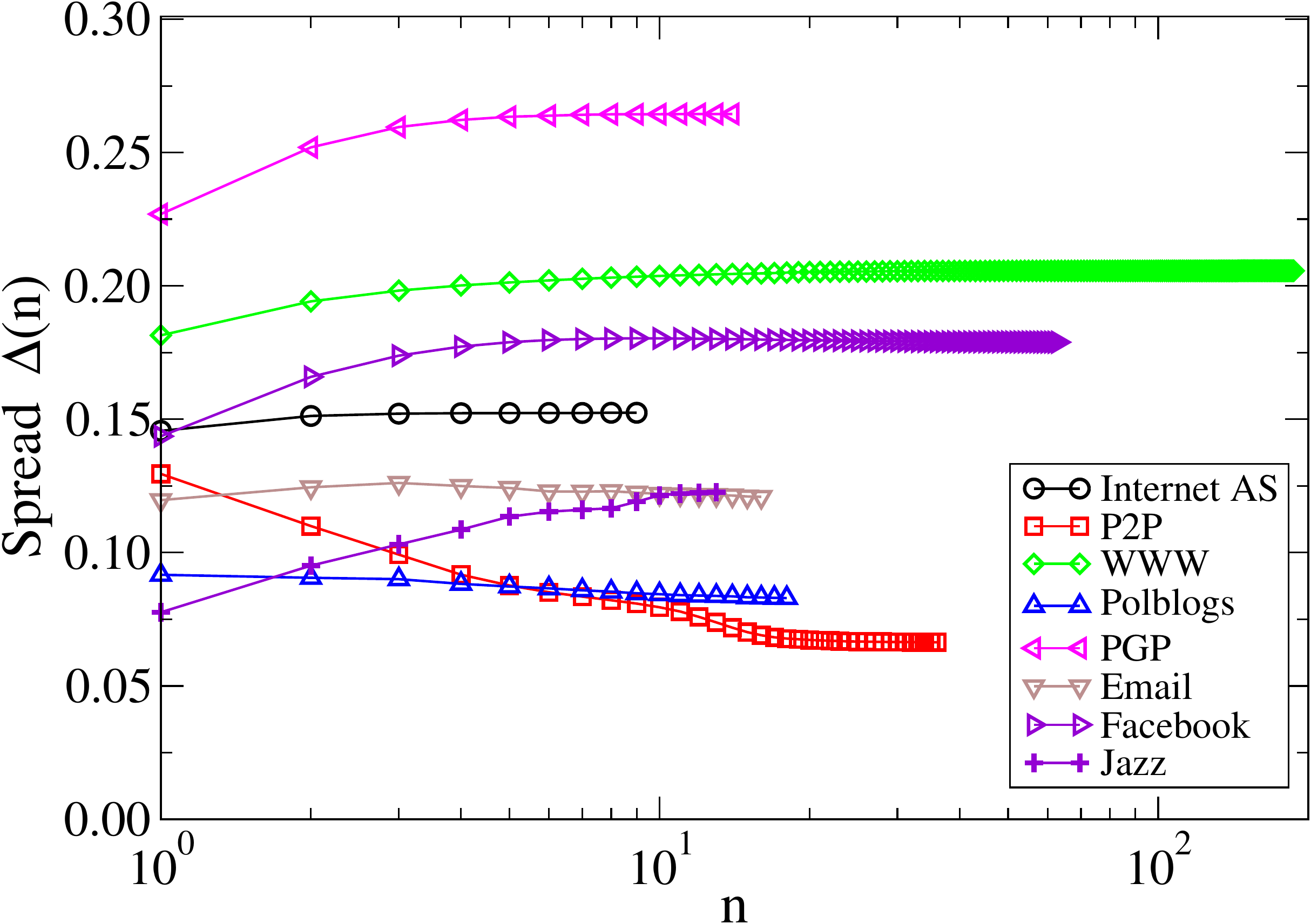}
  \caption{Spread of the $H(n)$ index over the average value
    $\overline{h^{(n)}}(q)$, as measured by the relative mean square
    spreading
    $\Delta(n) = (\sum_i[h^{(n)}_i/ \overline{h^{(n)}}(q_i) -
    1]^2/N)^{1/2}$, for real correlated networks.}
  \label{fig:Spread_h}
\end{figure}
The range of the spread changes when increasing the order $n$ of the
$H(n)$ index, as we can see in Fig.~\ref{fig:Spread_h}, where we compute
the relative mean square spreading
$\Delta(n) = (\sum_i[h^{(n)}_i/ \overline{h^{(n)}}(q_i) - 1]^2/N)^{1/2}$
for the real networks considered. In this figure we observe that, in
some cases (e.g. P2P), the $H(n)$ structure becomes more correlated with
degree for increasing order $n$, as reflected in a decreasing
$\Delta(n)$, while the opposite behavior is observed in other cases
(e.g. PGP, Facebook, and Jazz), which show a clearly increasing
$\Delta(n)$. In the rest of the networks, on the other hand, the spread
of the $H(n)$ structure is essentially independent of $n$. 

The effects of degree correlations can be gauged in
Fig.~\ref{fig:real1}, by comparing with the
$\overline{h^{(n)}}_\mathrm{ran}(q)$ computed from randomized versions
of the corresponding real networks. As we can see, again in some cases
(e.g. Polblogs, Email, Jazz, and to a lesser degree P2P and Facebook),
the actual average values of $\overline{h^{(n)}}(q)$ are almost
indistinguishable from the randomized, uncorrelated counterparts. From
inspection of Table~\ref{table_topology}, one can conclude that the
effects of correlations are apparently more strongly linked with the
network heterogeneity, as measured by the $\kappa$ parameter, than to
the actual value of the Pearson coefficient $r$: The larger $\kappa$,
the stronger the effects of degree correlations on the $H(n)$
topological strucuture.

\begin{figure}[t]
  \includegraphics[width=\columnwidth]{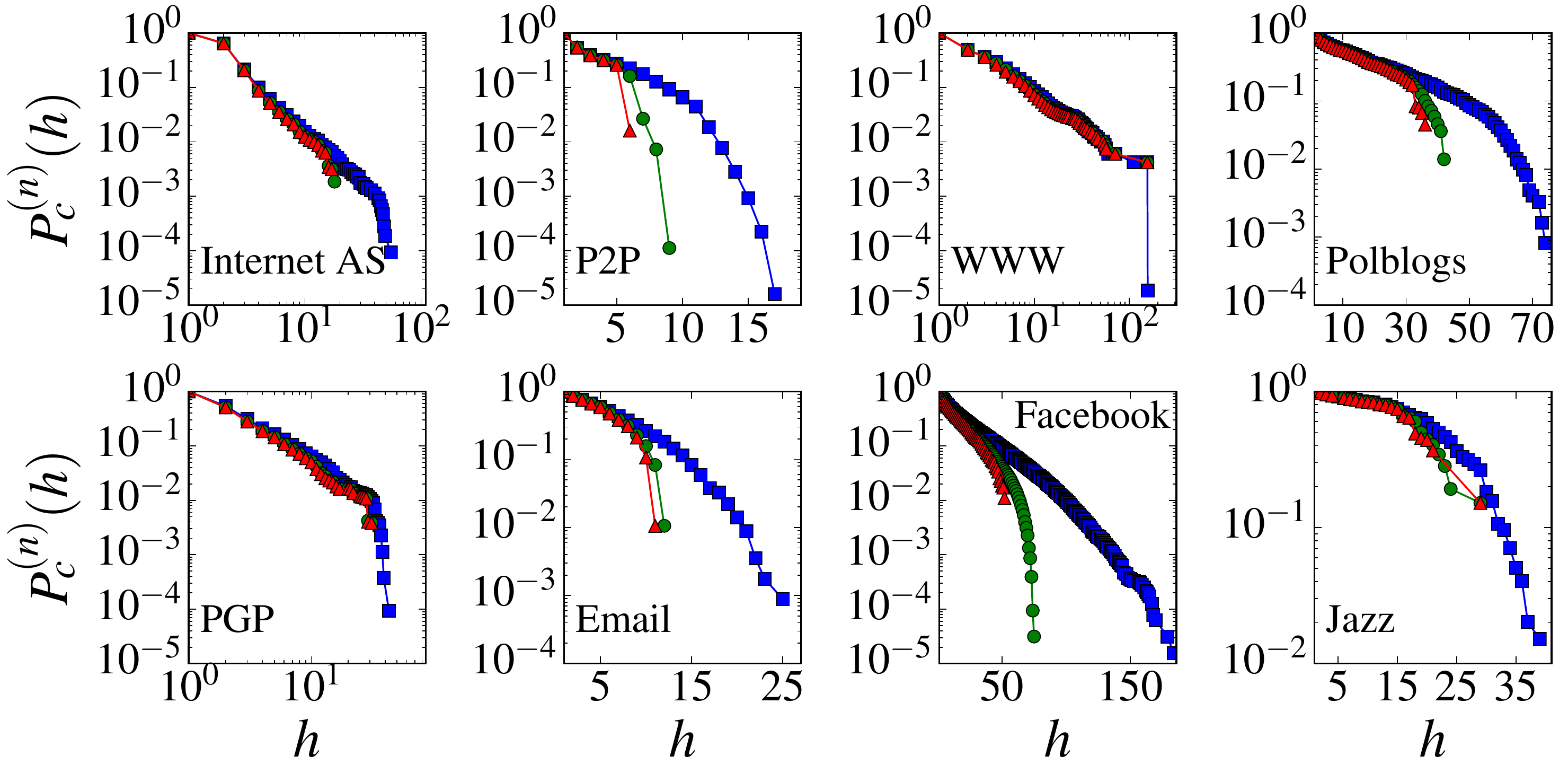}
  \caption{Cumulative distribution $P^{(n)}_c(h)$ of $H(n)$ indices for
    different values of $n$, computed from several real networks: $n=1$,
    squares; $n=2$, circles; $n=\infty$, triangles.}
  \label{fig:P_of_h_real}
\end{figure}

In Fig.~\ref{fig:P_of_h_real} we finally present the plots of cumulative
distributions of Hirsch indices,
$P^{(n)}_c(h) = \sum_{h' \geq h}P^{(n)}(h)$ (in other words, the size
distribution of the $H(n)$-cores), computed in real networks for
different values of $n$. In some cases, e.g. Internet AS, WWW and PGP, a
very clear power-law behavior, in agreement with the theoretical
prediction for uncorrelated networks, is observed. These three networks
are, in fact, the most heterogeneous among those considered, see
Table~\ref{table_topology}. The other networks, on the other hand, are
more compatible with an exponential $H(n)$ distribution, as expected
from the theory for essentially homogeneous networks. These observations
are again compatible with the prediction of a vanishing $K$-core
structure in homogeneous networks (power-law with $\gamma>3$). In the
case of highly heterogeneous real networks (Internet AS, WWW and PGP),
the $H(n)$ distribution is quite robust to changes in the order $n$. In
the particular case of the WWW, the $K$-core distribution ($n=\infty$)
is essentially equal to the distribution corresponding to $n=1$. For the
less heterogeneous networks, the $P^{(n)}_c(h)$ distributions become
narrower when increasing $n$. The possible exception is given by the
Jazz network which is, incidentally, the smallest one.

\section{The \hi as an indicator of influence}
\label{sec:test-hi-as}

In recent years a lot of activity has regarded the identification of
influential spreaders in networks~\cite{Klemm2012,kitsak2010,Lu16}, i.e.
the nodes which maximize the extent of spreading events initiated by
them. The goal is to find which of the many possible centralities based
solely on the network topology (such as degree, betweenness, $K$-core,
etc.) is most correlated with the actual spreding power of nodes.  In
Ref.~\cite{Lu2016} it is argued that the \hi $H(1)$ of a node is a
useful tool to quantify node influence for spreading phenomena, modeled
by the Susceptible-Infected-Removed (SIR) epidemic
dynamics~\cite{Pastor-Satorras:2014aa}.  This claim is based on the
presence of a maximum for $n=1$ in the plot of Kendall's $\tau$
coefficient as a function of the order $n$ of the $H(n)$ index
considered~\cite{Lu2016}.  Kendall's $\tau$ is a measure of the rank
correlation among two ordered sets and thus used to quantify the
agreement between the predicted influence and the one observed in
simulations.  The maximum for $n=1$ implies that the \hi is a better
indicator of influence than either degree ($n=0$) or $K-$core index
($n=\infty$).

However, Kendall's $\tau$ takes into account the ranking of all nodes in
the network, and therefore its value is strongly biased by the many
vertices which have very little influence and occupy middle and low
positions in the ranking.  This somehow washes out the effect of the
truly highly influential spreaders. In this sense, the imprecision
function $\epsilon(\rho)$ proposed in Ref.~\cite{kitsak2010} turns out
to be a more precise measure of the predictive power of centralities.
To define the imprecision function, consider a network of size $N$, and
a parameter $\rho$ in the range $0<\rho \leq 1$. Let us define
$S^{(x)}(\rho)$ as the set of the top $\rho N$ vertices according to the
rank given by some centrality measure $x$ and $S^{(\mathrm{SIR})}(\rho)$
as the set of the actual top $\rho N$ spreaders, as measured by means of
SIR simulations. This actual ranking of spreaders is based on the
average size of the outbreaks occurring when each node is a single
isolated seed.  The average outbreak size generated by the $\rho N$ most
highly ranked nodes according to the centrality measure $x$ is
\begin{equation}
  Z^{(x)}(\rho) = \frac{1}{N \rho} \sum_{i \in S^{(x)}(\rho)}
  \langle Q_i \rangle,
  \label{eq:zeta}
\end{equation}
where $\langle Q_i \rangle$ is the average size of outbreaks initiated
by node $i$, as measured by means of SIR numerical simulations.  If
$Z^{(\mathrm{SIR})}(\rho)$ is the analogous quantity of
Eq.~(\ref{eq:zeta}) but computed over the set
$S^{(\mathrm{SIR})}(\rho)$, the imprecision function corresponding to
centrality $x$ is defined as
\begin{equation}
  \epsilon^{(x)}(\rho) = 1 -
  \frac{Z^{(x)}(\rho)}{Z^{(\mathrm{SIR})}(\rho)}. 
  \label{eq:11}
\end{equation}
If the centrality $x$ perfectly identifies the most efficient spreaders,
then the imprecision function is equal to zero for every $\rho$.  A
large value of $\epsilon^{(x)} (\rho)$ indicates instead that
the centrality is not a good predictor of the spreading power of the
top $\rho N$ spreaders.

\begin{figure}[t]
  \includegraphics[width=\columnwidth]{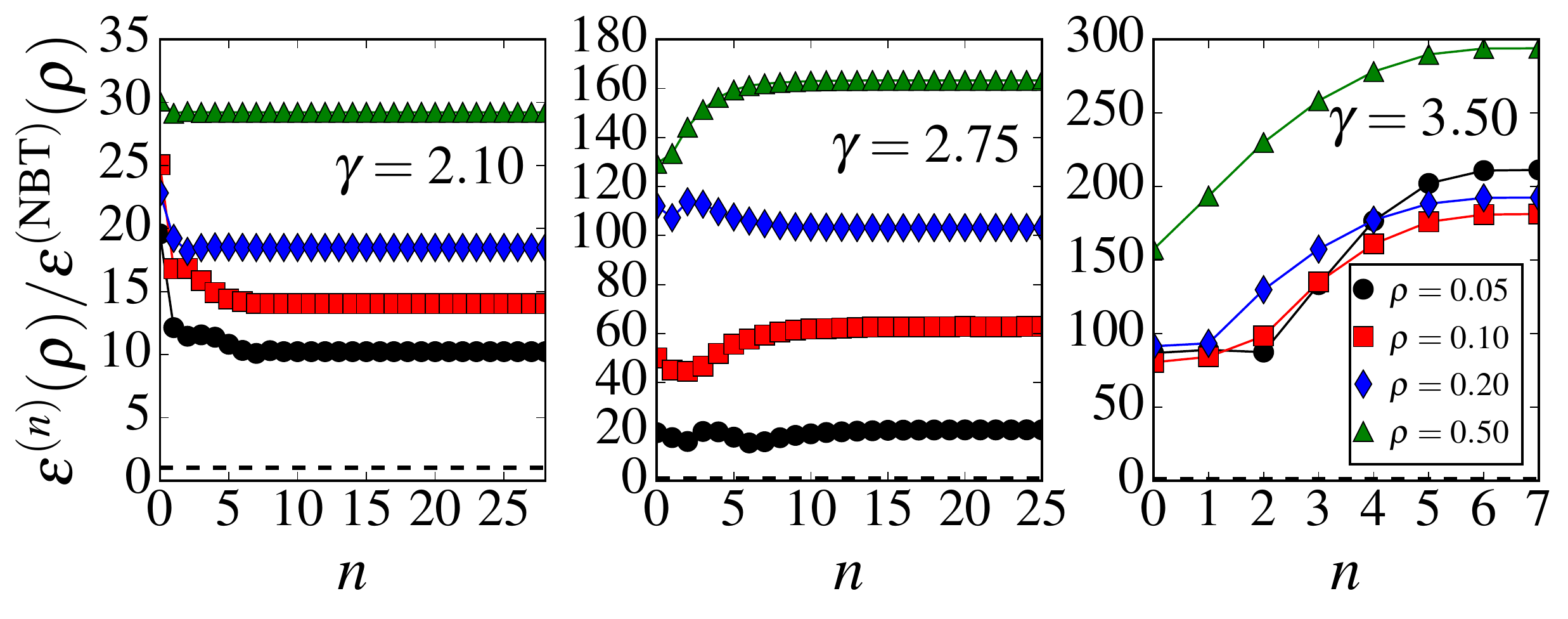}
  \caption{Plot of the imprecision function $\epsilon^{(n)}(\rho)$ for UCM
    networks with different degree exponent $\gamma$ and size $N=10^5$, as a function of
    the order $n$ of the generalized \hi, for several values of $\rho$. 
    The function $\epsilon^{(n)}(\rho)$ is
    normalized by the imprecision function corresponding to the NBT
    centrality.  
    The average outbreak size $\langle Q_i \rangle$ is computed over at
    least $10^4$ realizations.  The value of $\beta$ is the critical
    $\beta_c$ for which the susceptibility has a
    maximum~\cite{Castellano16}, namely $\beta_c = 0.0200$ for
    $\gamma = 2.10$, $\beta_c = 0.0650$ for $\gamma = 2.75$,
    $\beta_c = 0.2005$ for $\gamma = 3.50$.}
  \label{influenceUCM}
\end{figure}

We compute the imprecision function $\epsilon^{(n)}(\rho)$ for
various values of $\rho$ as a function of the order $n$ of the
generalized \hi used as centrality, for uncorrelated scale-free networks
generated using the UCM model (with various degree exponents $\gamma$). We
compare these quantities with the imprecision function calculated
using the non-backtracking (NBT) centrality \cite{2014arXiv1401.5093M},
$\epsilon^{(\mathrm{NBT})}(\rho)$ as an indicator of spreading influence
for fixed values of $\rho$. The NBT centrality has been recently proven
to be a very good predictor of influence for the SIR model
\cite{Radicchi2016}. For comparison, in Fig.~\ref{influenceUCM} we
present plots of the ratio
$\epsilon^{(n)}(\rho)/\epsilon^{(\mathrm{NBT})}(\rho)$, computed for UCM
networks of size $N=10^5$.

As we observe from Fig.~\ref{influenceUCM}, for small values of
$\gamma$, and large $\rho$, it appears that values of $n>0$ provide
a better estimate than the simple degree ($n=0$), which is reflected in
a smaller relative value of $\epsilon^{(n)}(\rho)$. The improvement of
$n>0$ over $n=0$ is superior for small values of $\rho$, indicating a
better performance in pinpointing the smallest set of most influential
spreaders.  This tendency appears to reverse for large degree exponents,
specially at $\gamma=3.5$. In this case, the effect can be understood by
the fact that random scale-free networks with exponent $\gamma>3$ do not
possess a $K$-core structure \cite{Dorogovtsev2006}: The iteration of
the \hi calculation leads thus to larger classes of nodes with the same
index, washing thus out the possibility of effective prediction. We
note, however, that for every $\gamma$ and $\rho$ considered, the
prediction of the NBT centrality is always better than the one for any
\hi $n$; the ratios plotted in Fig.~\ref{influenceUCM}
are always much larger than $1$ (marked as a dashed line). The performance
of the \hi as influence indicator steadily worsens when increasing
$\gamma$.

\begin{figure}[t]
  \includegraphics[width=\columnwidth]{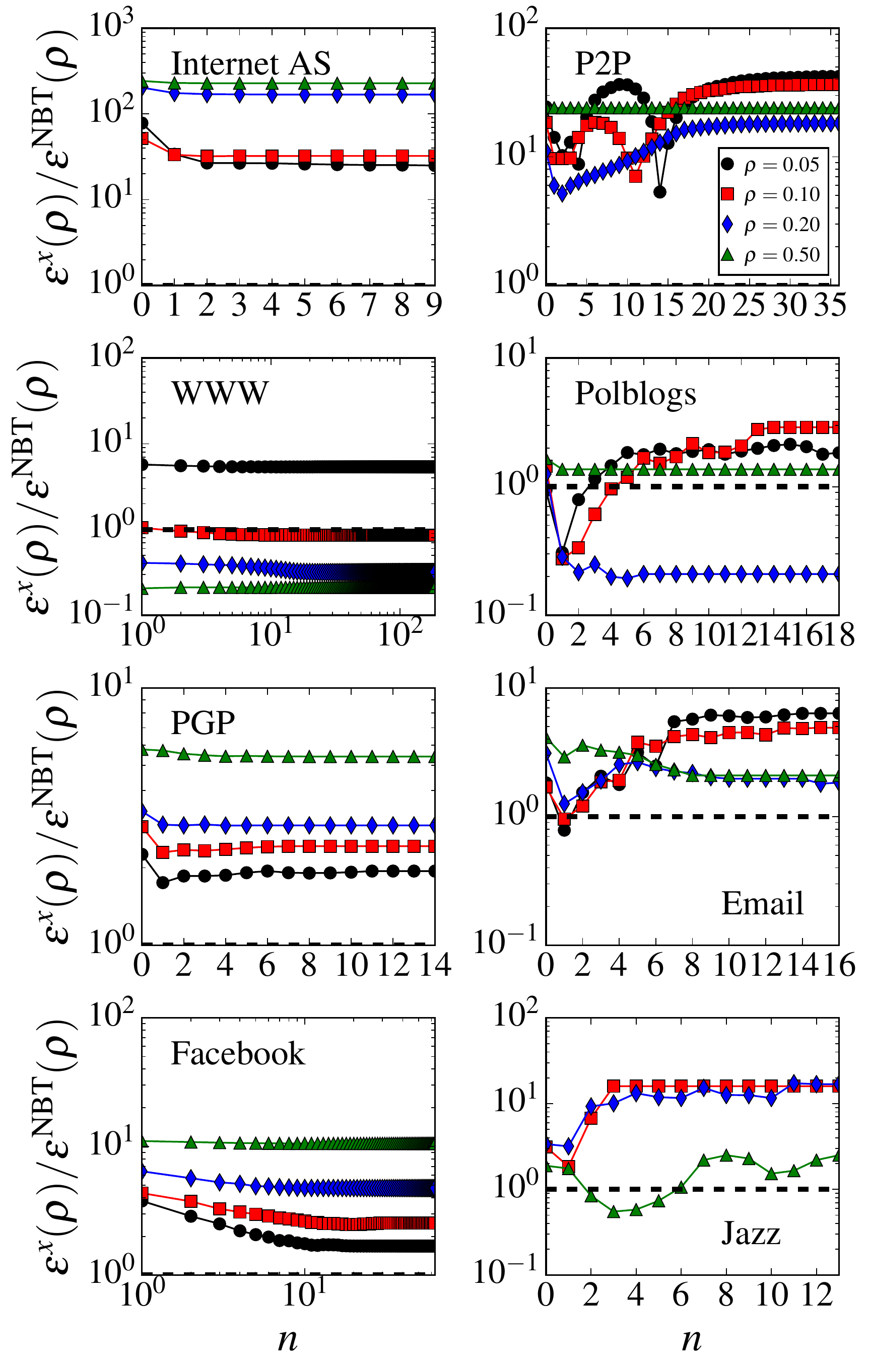}
  \caption{Plot of the imprecision function $\epsilon^{(n)}(\rho)$ for real
    networks, as a function of the order $n$ of the generalized \hi, for
    several values of $\rho$.
    The
    functions $\epsilon^{(n)}(\rho)$ are normalized by the imprecision
    function corresponding to the NBT centrality. 
    The average outbreak size $\langle Q_i \rangle$ is computed over at
    least $10^4$ realizations (up $10^6$ realizations for small
    networks).  The value of $\beta$ is the critical $\beta_c$ for which
    the susceptibility has a maximum~\cite{Castellano16}, see
    Table~\ref{table_topology}.}
  \label{influenceReal}
\end{figure}

\begin{figure}[t]
  \includegraphics[width=\columnwidth]{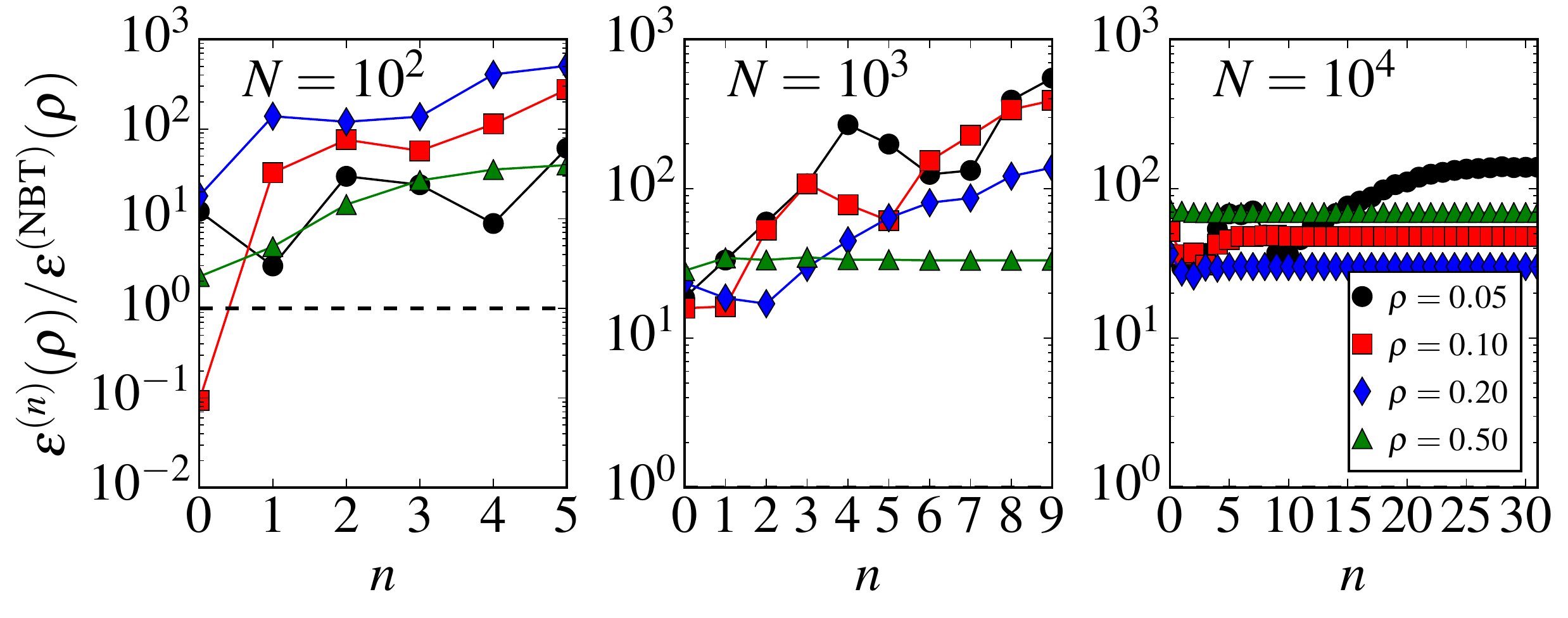}
  \caption{Plot of the imprecision function $\epsilon^{(n)}(\rho)$ for
    UCM networks with degree exponent $\gamma=2.1$ and different size,
    as a function of the order $n$ of the generalized \hi, for several
    values of $\rho$.  The function $\epsilon^{(n)}(\rho)$ is normalized
    by the imprecision function corresponding to the NBT centrality.
    The average outbreak size $\langle Q_i \rangle$ is computed over at
    least $10^4$ realizations.  The value of $\beta$ is the critical
    $\beta_c$ for which the susceptibility has a
    maximum~\cite{Castellano16}, namely $\beta_c = 0.275$ for $N=10^2$,
    $\beta_c = 0.115$ for $N=10^3$, $\beta_c = 0.046$ for $N=10^4$.}
  \label{influenceUCMSize}
\end{figure}

In Fig.~\ref{influenceReal} we present the same sort of analysis,
performed now for the set of 8 real correlated networks described in
Sec.~\ref{sec:correlated-networks}.  It turns out that in some cases (in
particular for small networks and small $\rho$) the imprecision function
has a minimum for $n=1$. This means that the Hirsch index $H(1)$
performs better than both the degree ($n=0$) and the $K$-core indicator
($n=\infty$) in agreement with the findings of
Ref.~\cite{Lu2016}. However the minimum (if present) is almost always
quite shallow, indicating that the improvement with respect to the
degree centrality is small. When comparing the efficiency of the \hi as
influence predictor with the NBT centrality, the latter performs usually
much better than degree, $K$-core or generalized $H(n)$ for any $n$, in
agreement with recent results in a more general
context~\cite{Radicchi2016}. Exceptions are the Polblogs, Email, Jazz
and WWW networks, in which apparently the \hi performs better than NBT
centrality for some values of $\rho$. The first three are the smallest
networks we consider, and thus it remains possible that this better
performance is due to finite size effects, as the performance of the NBT
centrality is guaranteed to be optimal only for uncorrelated nets of
infinite size. We explore further the
effects of small network size in Fig.~\ref{influenceUCMSize}, the
normalized imprecision function computed in uncorrelated UCM with degree
exponent $\gamma=2.1$ and different sizes $N$. As we can see, for the
smallest network size $N=10^2$, the performance of the $H(n)$ index is
close to the one of the NBT centrality, overcoming it for $n=0$ (the
degree) and $\rho=0.05$. It is however clear that, for increasing $N$,
the performance of the NBT centrality increases, becoming dominant in
the large network size limit.

The case of the WWW network is particular, due to the
peculiar structure of the NBT centrality it exhibits. As a matter of
fact, in this network the NBT centrality is densely localized in a set
of around $10000$ nodes, all the rest having a much smaller
centrality. This implies that the NBT centrality is a very good
predictor for small $\rho$, but for larger values one is mixing the
localization core with other irrelevant nodes, and the predictive power
strongly diminishes, being superseded by the \hi centrality.  We can
conclude therefore that generalized Hirsch indices for any $n$ are not
in general particularly useful tools for the identification of
influential spreaders.

\section{Conclusions}
\label{sec:conclusions}

In this paper, we have presented a detailed characterization of the
topological properties of the $H(n)$-shell structure introduced in
Ref.~\cite{Lu2016}, interpolating, as $n$ grows, between the degree and
the $K$-shell structure of the network.  A theoretical analysis performed
on uncorrelated networks within the annealed network approximation
reveals that the value of the \hi of order $n$ is strongly related with
degree, in the sense that nodes within a given $H(n)=h$-shell have a
well defined average degree with small fluctuations around it.  This
observation indicates that the value of the $H(n)$ index of a vertex is
strongly correlated with its degree. 
In uncorrelated power-law degree-distributed networks, it is possible to
derive in detail the exponents governing the dependence of the average
value of $H(n)$ on the degree $q$, as well as the decay of the
probability distribution of $H(n)$ indices. A numerical check confirms
the validity of the theoretical predictions.  The presence of
correlations and the quenched nature of the topology complicate the
picture in real-world networks.  The analysis of a limited set of
real-world topologies indicates nevertheless that the strong relation
between $H(n)$ index and degree often remains valid in average. In some
cases, however, the spread of the \hi for nodes of fixed degree $q$ is
larger than the prediction for uncorrelated networks.

Finally, we have shown that generalized $H(n)$ indices do not possess
any special property as indicators of influential spreaders in large
networks, where they are in general outperformed by the recently
proposed NBT centrality. Only for very small networks they can provide
in some cases better performance as influence indicators.  In
particular, the performance for $n=1$ is only slightly better than the
one of degree centrality.

To sum up our observations, the generalized $H(n)$ indices appear as an
elegant mathematical concept to connect the degree of a node with its
coreness. However, their practical application as a topological
observable in network science is diminished by the strong correlation
with degree observed in both uncorrelated networks and many instances of
real networks, as well as by their poor performance, compared to the
NBT-centrality, as a predictor of spreading influence in reasonably
large networks.

\begin{acknowledgments}
  We acknowledge financial support from the Spanish MINECO, under
  project FIS2013-47282-C2-2, and EC FET-Proactive Project MULTIPLEX
  (Grant No. 317532). R. P.-S. acknowledges additional financial support
  from ICREA Academia, funded by the Generalitat de Catalunya.
\end{acknowledgments}



%

\end{document}